\documentclass[submission,copyright,creativecommons]{eptcs}
\providecommand{\event}{FSFMA 2014} 

  \newcommand{\com}[1]{$\Leftarrow$ {Card1} $\Rightarrow$}

\newtheorem{definition}{Definition}
\newtheorem{theorem}{Theorem}

\newtheorem{proposition}{Proposition}

\usepackage{pgf}  
\usepackage{tikz} 
\usepackage{xcolor}
\usepackage{listings}
\usepackage{amsmath}
\usepackage{amssymb}
\usepackage{calc}
\usepackage{rotating}
\usepackage{subfigure}
\usepackage[ruled,vlined,linesnumbered]{algorithm2e}

\usepackage{pgf}
\newcommand{\todo}[1]{{\bf \textcolor{red}{[Card1]}}}

\lstdefinelanguage{pha}
{
morekeywords=[1]{automaton, contr_var, synclabs, loc, while, wait, when, sync, \
\
do, goto, end, initially},
alsoletter={=},
morekeywords=[2]{==,=},
keywordstyle=[2]{\tt},
sensitive,
morecomment=[l]{//},
morecomment=[s]{/*}{*/},
morestring=[b]'',
escapeinside={/*@}{@*/},
basicstyle=\sffamily\footnotesize,
breaklines=true,
xleftmargin=0.8cm,
numbers=left,
mathescape=true
}[keywords,comments,strings]

\lstdefinelanguage{imi}
{
morekeywords=[1]{automaton, var, clock, discrete, parameter, analog, synclabs, \
\
loc, while, wait, when, sync, do, goto, end, initially},
alsoletter={=},
morekeywords=[2]{==,=},
keywordstyle=[2]{\tt},
sensitive,
morecomment=[l]{//},
morecomment=[s]{/*}{*/},
morestring=[b]'',
scapeinside={(@}{@)},
basicstyle=\sffamily\footnotesize,
breaklines=true,
xleftmargin=0.8cm,
numbers=left,
mathescape=true
}[keywords,comments,strings]

\title{Correct-by-design Control Synthesis for  Multilevel Converters using State Space Decomposition} 

\author{Gilles Feld
\institute{SATIE, ENS Cachan \& CNRS, France}
\and Laurent Fribourg
\institute{LSV, ENS de Cachan \& CNRS, France}
\and Denis Labrousse
\institute{SATIE, ENS Cachan \& CNRS, France}
\and Bertrand Revol
\institute{SATIE, ENS Cachan \& CNRS, France}
\and Romain Soulat
\institute{LSV, ENS de Cachan \& CNRS, France}
}

\def\titlerunning{Correct-by-design control synthesis for multilevel converters}
\def\authorrunning{G. Feld, L. Fribourg, D. Labrousse, B. Revol  \& R. Soulat\
}

\begin{document}

\maketitle

\begin{abstract} 
High-power converters based on elementary switching cells are more and 
more used in the industry of
power electronics owing to various advantages such as lower voltage 
stress and reduced power loss.
However, the complexity of controlling such converters is a major 
challenge that 
the power manufacturing industry has to face with. The synthesis of 
industrial  switching controllers
relies today on heuristic rules and empiric simulation. The
state of the system is not guaranteed to stay within
the limits that are admissible for its correct electrical behavior.
We show here how to apply a formal method in order
to synthesize a correct-by-design control that
guarantees that the power
converter will always stay within a predefined safe zone of 
variations for its input parameters.
The method is applied in order to synthesize a
correct-by-design control for 5-level and 7-level power converters 
with a flying capacitor topology.
 We check the validity of our approach
by numerical simulations for 5 and~7 levels. We also perform physical experimentations
using a prototype built by SATIE laboratory for 5 levels.
\end{abstract}

\section{Introduction}
Switched control has gained much attention recently
due to its property of being easily 
implemented,
especially in the field of power converters.
Power converters play an important role in the field of renewable energy:
they are used to connect renewable sources to powergrids,
optimize the efficiency of solar panels
and  wind generators (see, e.g., \cite{CPM09}). 
In some topologies, there is however a dramatic increase
of the number of switches, which entails an increasing number of degrees 
of freedom, and complicates the controller design
. There is therefore a niche of application for formal methods
in order to produce correct-by-design control methods.
The general function of a multilevel power converter is to
synthesize a desired voltage from several levels of DC
voltage. For this reason, multilevel power converters
can easily provide the high power required by large
electric drive systems. 
A
multilevel converter is 
a power converter
made of  capacitors and
switching cells (as well as opposite switching
cells which are in complementary positions);

\begin{figure}[!ht]
\centering
\includegraphics[scale= 0.7]{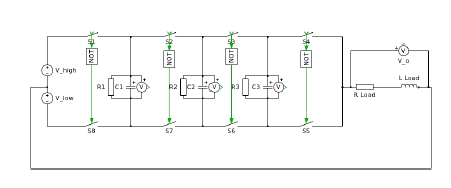}
\caption{Electrical scheme of a $5$-level flying capacitor converter}
\label{fig:scheme_5}
\end{figure}
\begin{figure}[!ht]
\centering
\includegraphics[scale= 0.3]{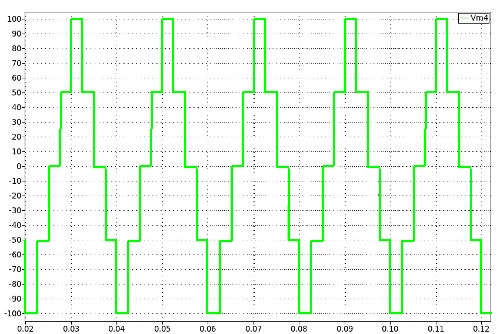}
\caption{Staircase output voltage waveform for a 5-level converter}
\label{fig:staircase_5}
\end{figure}

In this paper,
we consider the design of control policies for power converters with 
a number of levels $\ell=5$ and  $\ell=7$. 
A multilevel converter for $\ell=5$ is schematized on 
Figure~\ref{fig:scheme_5}.  
According to the positions of the cells,
one is able to
fraction the load voltage. By controlling the global position
of the switches during a simple fixed time-stepping
procedure, it is then possible to generate a staircase voltage
with levels that approximates a triangular or a sinusoidal waveform (see
Figure \ref{fig:staircase_5}, for $5$ levels).

%

The problem which arises is to select the appropriate switching
control strategy among a number of combinations of
switch positions which increases exponentially with the number
of levels (and pairs of switches). A crucial 
difficulty comes from the fact that, in order to be admissible,
the control of the switching cells must guarantee that the
voltages across the cell-capacitors are constrained within a
certain range defined by the device blocking voltage rating.
The control must thus guarantee a {\em safety property}, called
``capacitor voltage balancing'': the voltage of each individual
capacitor should stay inside a limited predefined interval.
The synthesis of 
industrial  switching controllers
relies today on heuristic rules and empiric simulation. The
state of the system is not guaranteed to always satisfy
capacitor voltage balancing.
In this paper, we show how to
synthesize a control,
 by applying 
a formal method, called
{\em state space decomposition
procedure} \cite{FS-rp13}.
The synthesized control is ``correct-by-design'' because it is ensured
to make the electrical state parameters 
of the system
stay within predefined safe zones of variations.
Numerical simulations, performed at levels $\ell=5,7$,
confirm the safety properties of the synthesized control.
Physical experimentations are also successfully performed
on a  prototype built by SATIE Electronics Laboratory,
at level $\ell=5$.\\

{\bf Outline of the paper}

In Section \ref{sec:decomp}, we present the principles
of the state space decomposition method.
In Section \ref{sec:multilevel}, 
we apply the method  in order to synthesize the control
of multilevel converters with a flying capacitor topology,
for 5 and~7 levels.
In Section \ref{sec:physical_exp}, we present
physical experimentations done with a prototype 
of 5-level converter. 
We conclude in Section \ref{sec:conclusion}.

\section{State Space Decomposition Method}\label{sec:decomp}
A multilevel converter can be seen as a ``switched system'',
where the different operating modes depend on the positions
of the switching cells.
In this section, we describe a general method
that is useful for proving properties of switched systems.
The method will be subsequently applied 
to multilevel converters in Section \ref{sec:multilevel}.
\subsection{Model of affine sampled switched systems}

A {\em switched system} $\Sigma$ is defined by a finite family of differential equations of the form $\{{\dot x} = f_u(x)\}_{u \in U}$ where $U$ is a finite 
set of {\em modes}
(see, e.g., \cite{GPT10,T09}).
In the following, we consider that the dynamics of the subsystems are {\em affine}
 (i.e., $f_u(x)$ is of the form $A_ux + b_u$ with $A_u\in \mathbb{R}^{n\times n}$ and $b_u$ a vector
of $\mathbb{R}^n$).
The control problem for a switched system $\Sigma$ is to find a piecewise constant law ${\bf u} : \mathbb{R}_{\geq 0} \rightarrow U$ 
in order to achieve some pertained goals.
The {\em switching instants} are the times at which ${\bf u}$ changes its value. An {\em affine sampled switched system}
 is a switched system for which the switching instants
occur at integer multiples of $\tau$ (called {\em sampling parameter}).
We will use ${\bf x}(t,x,u)$ to denote the point reached by $\Sigma$ at time
$t$ under mode $u$ from the initial condition $x$.
This gives a  transition relation $\rightarrow_u^{\tau}$ defined for $x$ and $x'$ in $\mathbb{R}^n$ by:
$x\rightarrow_u^{\tau} x' \mbox{ iff } \mathbf{x}(\tau,x,u)=x'$.
Given a set $X\subset\mathbb{R}^n$, we define:
$$Post_u(X)=\{x' \ |\ x\rightarrow_u^{\tau} x'  \mbox{ for some $x\in X$}\}.$$
It can be seen that $Post_u(X)$ is the result of an affine transformation of the form
$C_uX + d_u$ with $C_u\in \mathbb{R}^{n\times n}$ and $d_u$ a vector of $\mathbb{R}^n$.

A {\em pattern} $\pi$ is defined as a finite sequence of modes. A 
{\em $k$-pattern} is a pattern of length at most $k$.
%
The mapping $Post_{\pi}$ is itself an affine transformation.

Given a pattern $\pi$ of the form $(u_1\cdots u_m)$, and a set $X\subset\mathbb{R}^n$,
the {\em unfolding of $X$ via $\pi$}\index{unfolding}, 
denoted by $\mathit{Unf}_{\pi}(X)$, 
is the set $\bigcup_{i=0}^mX_i$ with:
\begin{itemize}
\item$X_0=X$,
\item $X_{i+1}=Post_{u_{i+1}}(X_i)$, for all $0\leq i\leq m-1$.
\end{itemize}
The unfolding thus corresponds to the set of all the 
intermediate states produced when applying pattern $\pi$ to 
the states of $X$.
\subsection{Safety control problem}
A safety property is typically expressed 
using a subset $S$ of the continuous state
space, called {\em safe set}.
In a simple formulation, $S$ is a {\em box},
i.e., a cartesian product of intervals that   
specify the minimum and
maximum values tolerated for each state component.
Given a safe set $S$, and a domain of interest $R\subseteq S$,
we can define the notion of ``safe control'' in this context as
follows.
\begin{definition}
Given  a domain of interest $R$ and safe set $S$ with $R\subseteq S$,
a {\em safe control of $R$ w.r.t. $S$} is a function that
associates to each $x\in R$ a pattern $\pi$ such that:
\begin{itemize}
\item $Post_{\pi}(\{x\})\subseteq R$, and
\item $Unf_{\pi}(\{x\})\subseteq S$. 
\end{itemize}
\end{definition}
Given a domain of interest $R$ and a set $S$ with $R\subseteq S$,
the safety control problem consists in finding
a safe control of $R$ w.r.t. $S$.
In \cite{FS-rp13},
in order to solve such a problem, we introduced
the notion of ``(safe) decomposition''.
%
\begin{definition}
Given a set $R\subset\mathbb{R}^n$
and a set $S$ with $R\subseteq S$, a  
{\em safe decomposition of $R$ w.r.t. $S$}
is a set $\Delta$
of the form $\{V_i,\pi_i\}_{i \in I}$, where~$I$ is a finite set of indices, $V_i$s are subsets of $R$, $\pi_i$s are $k$-patterns, such that:
\begin{itemize}
\item $\bigcup_{i \in I}V_i = R$, 
\item for all $i \in I$: $Post_{\pi_i}(V_i) \subseteq R$, and
\item for all $i\in I$: $Unf_{\pi_i}(V_i)\subseteq S$.
\end{itemize}
\end{definition}



A decomposition $\Delta=\{(V_i,\pi_i)\}_{i\in I}$ naturally
induces a  {\em state-dependent control} on $R$.
Furthermore, the controlled trajectories starting from
$R$ never leave~$S$.
%
Indeed, given a starting state $x_0$ in $R$,
we know that $x_0\in V_i$ for some $i\in I$
(since $R=\bigcup_{i\in I} V_i$); one thus applies $\mathrm{\pi}_i$
to $x_0$, which gives a new state $x_1$ that belongs itself to $R$
(since $Post_{\pi_i}(V_i) \subseteq R$); 
furthermore, since 
$Unf_{\pi_i}(V_i)\subseteq S$,
all the intermediate states produced by application of $\pi_i$
are guaranteed to belong to~$S$. 
The process can then be
repeated on $x_1$, and so on iteratively.
Formally, we have:
\begin{proposition}\label{prop:safe}
Suppose that $\Delta$
is a safe decomposition of $R$ w.r.t. $S$. Then the
control of $R$ induced by $\Delta$ is safe w.r.t $S$.
\end{proposition}
%
The problem of finding a safety controller thus reduces
to the problem of finding
a safe decomposition $\Delta$.
The latter problem can be solved by using
the {\em state space decomposition method} \cite{FS-rp13},
as explained below.
%

\subsection{State space decomposition method}


We give here a simple algorithm,
adapted from \cite{FS-rp13},
called Decomposition algorithm.  Given a set $R$ and a set $S$
with $R\subseteq S$, the algorithm
outputs, when it succeeds, a decomposition $\Delta$ of $R$ w.r.t $S$,
of the form $\{V_i,\pi_i\}_{i \in I}$.
The input sets $R$ and $S$ are given under the form of {\em boxes} of $\mathbb{R}^n$ (i.e.,  cartesian products of $n$ closed intervals).
The subsets $V_i$s of $R$ are boxes that are obtained by 
repeated bisection.
At the beginning,
the Decomposition procedure calls sub-procedure Find$\_$Pattern in order
to get a $k$-pattern $\pi$ such that $Post_{\pi}(R)\subseteq R$
and $Unf_{\pi}(R)\subseteq S$. 
If it succeeds, then it is done.
Otherwise, it divides $R$ 
into $2^n$ 
sub-boxes $V_1,\dots,V_{2^n}$ of equal size.
If for each $V_i$, Find$\_$Pattern 
gets a $k$-pattern $\pi_i$ such that $Post_{\pi_i}(V_i)\subseteq R$
and $Unf_{\pi_i}(V_i)\subseteq S$,
it is done.
If, for some $V_j$, no such pattern exists,
the procedure is recursively applied to $V_j$.
It ends with success when a safe decomposition 
of $R$ w.r.t. $S$ is found,
or failure when
the maximal degree $d$ of decomposition is reached.
The algorithmic form of the procedure is given in Algorithms \ref{alg:decompo} and \ref{alg:findPat}.
(For the sake of simplicity, we consider the case 
of dimension $n=2$, but the extension to $n>2$ is straightforward.)
The main procedure  
Decomposition($W,R,S,D,K$)
is called with $R$ as input value for~$W$, $d$
for input value for $D$, and $k$ as input value for $K$;
it returns either $\langle\{(V_i,\pi_i)\}_{i},True\rangle$ with $\bigcup_i V_i=W$,
$\bigcup_i Post_{\pi_i}(V_i)\subseteq R$,
$\bigcup_i Unf_{\pi_i}(V_i)\subseteq S$
or $\langle\_ ,False\rangle$.
Procedure Find$\_$Pattern($W$,$R$,$S$,$K$) looks
for a $K$-pattern  $\pi$ for which
$Post_{\pi}(W)\subseteq R$ and $Unf_{\pi}(W)\subseteq S$:
it  selects all the $K$-patterns 
by non-decreasing length order
until either it finds such a pattern~$\pi$ 
(output: $\langle\pi, True\rangle$), or
none exists
(output: $\langle\_, False\rangle$).
The correctness of the procedure is stated as follows.

\begin{theorem}\label{th:procedure}
If Decomposition($R$,$R$,$S$,$d$,$k$) 
returns $\langle\Delta, True\rangle$, then $\Delta$ is a safe
decomposition of $R$ w.r.t. $S$.
\end{theorem}

\begin{algorithm}[!ht]
 \KwIn{A box $W$, a box $R$, a box $S$, a degree $D$ of decomposition, 
a length $K$ of pattern}
  \KwOut{$\langle\{(V_i,\pi_i)\}_{i},True\rangle$ with $\bigcup_i V_i=W$,
$\bigcup_i Post_{\pi_i}(V_i)\subseteq R$ and $\bigcup_i Unf_{\pi_i}(V_i)\subseteq S$
or $\langle\_ ,False\rangle$}
  $(\pi,b) := Find\_Pattern(W,R,S,K)$\\
  \If{$b=True$}{
    \Return{$\langle\{(W,\pi)\},True\rangle$}
  }
  \Else{
      \If{$D=0$}{
	  \Return{$\langle\_ ,False\rangle$}
       }
       \Else{Divide equally $W$ into $(W_1,\cdots, W_{2^{n-2}})$ \ \ \ \\
	    \For{$i=1\dots2^{n-2}$}{
	      $(\Delta_i,b_i)$ := Decomposition($W_i$,$R$,$S$,$D - 1$,$K$)\\
	    }
 	      \Return{$(\bigcup_{i=1\dots 2^{n-2}} \Delta_i,\bigwedge_{i=1\dots 2^{n-2}} b_i)$}
 	    }
}
\caption{Decomposition($W,R,S,D,K$)}  
\label{alg:decompo}  
\end{algorithm}

 \begin{algorithm}[!ht]
 \KwIn{A box $W$, a box $R$, a box $S$ a length $K$ of pattern}
 \KwOut{$\langle\pi,True\rangle$ with $Post_{\pi}(W)\subseteq R$ and $Unf_{\pi}(W)\subseteq S$, or $\langle\_, False\rangle$ when 
no pattern maps $W$ into $R$}
   \For{$i=1\dots K$}{$\Pi :=$ set of patterns of length $i$\\
    \While{$\Pi$ is non empty}{Select $\pi$ in $\Pi$\\ $\Pi:= \Pi\setminus  \{\pi\}$\\ \If{$Post_{\pi}(W) \subseteq R$ and $Unf_{\pi}(W) \subseteq S$}
    {\Return{$\langle\pi,True\rangle$}}}
    }
   \Return{$\langle\_ ,False\rangle$}
 \caption{Find\_Pattern($W,R,K$)}
\label{alg:findPat}
 \end{algorithm}

In \cite{FS-rp13},
we have developed a tool
that implements the Decomposition procedure, using zonotopes \cite{K98},
and is written in Octave \cite{octave}. 
We now describe the application of this tool,
called MINIMATOR \cite{minimator}, for synthesizing controllers
of multilevel converters.

\section{Application to Multilevel Converters}\label{sec:multilevel}
\subsection{Multilevel converters as switched systems}
There are different possible topologies
for multilevel power converters: neutral-point clamped,
cascaded H-bridge,  Modular Multilevel Converter (see e.g., \cite{SMVSSDSB12,DTOC09,LM03,MF92}). We focus
here on the flying capacitor topology \cite{MF92}. 
The electrical scheme of a 5-level converter 
was given in Figure \ref{fig:scheme_5}.
%
There are  4 pairs of switching cells 
$S_1, S_2, S_3, S_4$ 
(the
high-side switch conducting position is indicated by 1 and
the lowside switch conducting position by 0), 
and 3 capacitors $C_1,C_2, C_3$.
The state of the system is $x(t)=[v_1(t)\ v_2(t)\ v_3(t)\ i(t)]^T$
where $v_j(t)$ is the voltage across $C_j$ ($1\leq j\leq 3$) and
$i(t)$ is the current flowing in the circuit.
The duration
of a cycle is  $T=8\tau$.
The {\em mode} of the system is characterized by the value ($0$ or $1$) 
of the switching cells, i.e., by the
value of vector $S = [S_1\ S_2\ S_3\ S_4]^T$.\footnote{Besides, we have: $S_5 = \lnot S_1$,
$S_6 = \lnot S_2$,
$S_7 = \lnot S_3$ and $S_8 = \lnot S_4$.}
There are thus $2^4=16$ modes.
A mode $S$
induces an output voltage $v_o$ of value 
$\Sigma_{j=1}^3 (S_{j+1}-S_j)v_j + S_1 v_{high} - (1-S_1) v_{low}$,
where $v_{low}$ and $v_{high}$ are the input voltages
of low level and high level respectively. 
For the sake of simplicity, we suppose:
$v_{high} = v_{low} = v_{input}$.
The system then outputs 5 different levels of voltage which
go from $-v_{input}$ up to $+v_{input}$ 
with steps at 
$-\frac{v_{input}}{2},0,\frac{v_{input}}{2}$.
The ideal value  $v_i^*$ of the voltage across
capacitor $C_i$ ($1\leq i\leq 3$) depends on the values of $v_{input}$. Here we use:
$v_{input}=100V$, and $v_1^*=150V$, $v_2^*=100V$, $v_3^*=50V$.
The 5-level converter can be seen as a switched system.
Given a mode $S$, the associated dynamics
is of the form $\dot{x}(t)=A_Sx(t)+b_S$ with:
%
$$A_{S} = \begin{pmatrix}
 -\frac{1}{R_1 C_1} & 0 & 0 & \frac{S_1 - S_2}{C_1} \\
 0 & -\frac{1}{R_2 C_2} & 0 & \frac{S_2 - S_3}{C_2} \\
 0 & 0 & -\frac{1}{R_3 C_3} & \frac{S_3 - S_4}{C_3} \\
 \frac{S_2 - S_1}{L_{Load}} & \frac{S_3 - S_2}{L_{Load}} &  \frac{S_4 - S_3}{L_{Load}} & -\frac{R_{Load}}{L_{Load}}
  \end{pmatrix}
\mathrm{and} \ b_{S} =
\begin{pmatrix}
 0 \\ 0 \\ 0 \\ \frac{(2S_1-1)v_{input}}{L_{Load}}
\end{pmatrix}
$$

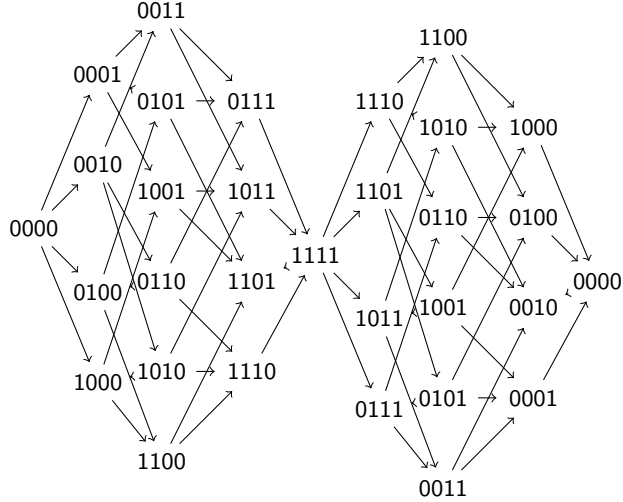
\begin{figure}[!ht]
 \centering
\begin{tikzpicture}[->,shorten >=1pt,auto,node distance=1.2cm,main
node/.style={font=\sffamily\footnotesize}]

  \node[main node] (1) {0000};
  \node[main node] [above right of =1] (2) {0010};
  \node[main node] [below right of = 1] (3) {0100};
  \node[main node] [below of = 3] (4) {1000};
  \node[main node] [above of = 2] (5) {0001};

  \node[main node] [above right of = 5] (6) {0011};
  \node[main node] [below of = 6] (7) {0101};
  \node[main node] [below of = 7] (8) {1001};
  \node[main node] [below of = 8] (9) {0110};
  \node[main node] [below of = 9] (10) {1010};
  \node[main node] [below of = 10] (11) {1100};

  \node[main node] [right of = 7] (12) {0111};
  \node[main node] [below of = 12] (13) {1011};
  \node[main node] [below of = 13] (14) {1101};
  \node[main node] [below of = 14] (15) {1110};
  \node[main node] [below right of = 13] (16) {1111};

  \node[main node] [above right of =16] (17) {1101};
  \node[main node] [below right of = 16] (18) {1011};
  \node[main node] [below of = 18] (19) {0111};
  \node[main node] [above of = 17] (20) {1110};

  \node[main node] [above right of = 20] (21) {1100};
  \node[main node] [below of = 21] (22) {1010};
  \node[main node] [below of = 22] (23) {0110};
  \node[main node] [below of = 23] (24) {1001};
  \node[main node] [below of = 24] (25) {0101};
  \node[main node] [below of = 25] (26) {0011};

  \node[main node] [right of = 22] (27) {1000};
  \node[main node] [below of = 27] (28) {0100};
  \node[main node] [below of = 28] (29) {0010};
  \node[main node] [below of = 29] (30) {0001};

  \node[main node] [below right of = 28] (31) {0000};

   \path[every node/.style={font=\sffamily\small}]
      (1) edge (2)
	  edge (3)
	  edge (4)
	  edge (5)
      (2) edge (6)
	  edge (9)
	  edge (10)
      (3) edge (7)
	  edge (9)
	  edge (11)
      (4) edge (8)
	  edge (10)
	  edge (11)
      (5) edge (6)
	  edge (7)
	  edge (8)
      (6) edge (12)
	  edge (13)
      (7) edge (12)
	  edge (14)
      (8) edge (13)
	  edge (14)
      (9) edge (12)
	  edge (15)
      (10) edge (13)
	  edge (15)
      (11) edge (14)
	  edge (15)
      (12) edge (16)
      (13) edge (16)
      (14) edge (16)
      (15) edge (16)

      (16) edge (17)
	  edge (18)
	  edge (19)
	  edge (20)
      (17) edge (21)
	  edge (24)
	  edge (25)
      (18) edge (22)
	  edge (24)
	  edge (26)
      (19) edge (23)
	  edge (25)
	  edge (26)
      (20) edge (21)
	  edge (22)
	  edge (23)
      (21) edge (27)
	  edge (28)
      (22) edge (27)
	  edge (29)
      (23) edge (28)
	  edge (29)
      (24) edge (27)
	  edge (30)
      (25) edge (28)
	  edge (30)
      (26) edge (29)
	  edge (30)
      (27) edge (31)
      (28) edge (31)
      (29) edge (31)
      (30) edge (31)

;

\end{tikzpicture}
\caption{Transition graph corresponding to a cycle of 5-level staircase
signal}
\label{fig:graph5}
\end{figure}

By controlling the modes at each sampling time, one can synthesize a
5-level staircase function. 
Not all the transitions between modes are admissible: we allow to switch only
one (pair of) cell(s) at a time.
The graph of admissible transitions during a cycle  is 
depicted in Figure \ref{fig:graph5}.
The nodes of the graph are labeled by the modes.
Each path represents a possible pattern for one cycle,
leading from voltage $-v_{input}$ (mode $0000$) to voltage $+v_{input}$ (mode $1111$) through voltages
 $-\frac{v_{input}}{2}$, $0$, $\frac{v_{input}}{2}$ then back to voltage $-v_{input}$ (mode $0000$)
through voltages $\frac{v_{input}}{2}$, $0$, $\frac{v_{input}}{2}$.
There are thus 576 possible patterns for generating a 5-level staircase signal on one cycle.

We explain in the following how to apply the tool MINIMATOR
in order to find a safe decomposition involving these patterns.

\subsection{Application of the Decomposition procedure to a 5-level converter}\label{sec:5-level}
 
We consider the following numerical values of the electrical parameters: 
$v_{input}= 100$V,
$R_{Load} = 50 \Omega$,
$C_1=C_2 = C_3 =  0.0012$F,
$L_{Load}= 0.2$H,
$R_1 = R_2 = R_3 = 20,000 \Omega$,
$T=8\tau=0.02s$ (which corresponds to a frequency of $50$Hz).

In this context, a 5-level converter outputs ideally
a staircase waveform with an amplitude of $200$V, centered around $0$V.
We consider that a variation of $\pm5V$ is admissible as it represents a variation of $10\%$ 
on the least charged capacitor~$C_3$.
It is interesting to notice that at each beginning of a cycle the value of $i$ is
null. This suggest to look for a state-dependent control which depends
only on the capacitor voltages $v_1,v_2,v_2$, and not on the value of $i$.
We will thus focus on the voltage dimensions of 
the control box $R$ and disregard its intensity dimension.
For $R$, we take
$R = [145,155] \times [95,105] \times [45,55]$,
which corresponds to a product of intervals centered around the
ideal values with a variation of $\pm 5V$ (i.e., $10\%$ 
of the least charged capacitor $C_3$).
For $S$, we take $R+\varepsilon$ with $\varepsilon=1V$,
which means that we have an additional tolerance of $\pm 1V$ for the fluctuations
occurring between two beginnings of cycle.

Given $R = [145,155] \times [95,105] \times [45,55]$ 
and $S = [144,156] \times [94,106] \times [44,56]$, we perform
the procedure of Decomposition, implemented in MINIMATOR tool,
on a machine equipped with an Intel core2 CPU X6800 at 2.93GHz and with
2GiB of Ram memory.
With parameters $d=1$ and $k=8$, the procedure outputs 
in 60 seconds a decomposition $\Delta=\{(V_i,\pi_i)\}_{i=1,\dots,8}$
with:
\begin{itemize}
 \item  $V_1 = [145,150] \times [95,100] \times [45,50]$
 \item  $V_2 = [145,150] \times [95,100] \times [50,55]$
 \item  $V_3 = [145,150] \times [100,105] \times [45,50]$
 \item  $V_4 = [145,150] \times [100,105] \times [50,55]$
 \item  $V_5 = [150,155] \times [95,100] \times [45,50]$
 \item  $V_6 = [150,155] \times [95,100] \times [50,55]$
 \item  $V_7 = [150,155] \times [100,105] \times [45,50]$
 \item  $V_8 = [150,155] \times [100,105] \times [50,55]$
\end{itemize}
and
\begin{itemize}
 \item $\pi_1$: $(0000\rightarrow 0001\rightarrow 0101\rightarrow 1101\rightarrow 1111\rightarrow 1101\rightarrow 0101\rightarrow 0001\rightarrow 0000)$ 
 \item $\pi_2$: $(0000\rightarrow 0100\rightarrow 0101\rightarrow 1101\rightarrow 1111\rightarrow 1101\rightarrow 0101\rightarrow 0100\rightarrow 0000)$ 
 \item $\pi_3$: $(0000\rightarrow 0001\rightarrow 0011\rightarrow 1011\rightarrow 1111\rightarrow 1011\rightarrow 0011\rightarrow 0001\rightarrow 0000)$ 
 \item $\pi_4$: $(0000\rightarrow 0010\rightarrow 0011\rightarrow 1011\rightarrow 1111\rightarrow 1011\rightarrow 0011\rightarrow 0010\rightarrow 0000)$ 
 \item $\pi_5$: $(0000\rightarrow 1000\rightarrow 1010\rightarrow 1110\rightarrow 1111\rightarrow 1110\rightarrow 1010\rightarrow 1000\rightarrow 0000)$ 
 \item $\pi_6$: $(0000\rightarrow 1000\rightarrow 1100\rightarrow 1101\rightarrow 1111\rightarrow 1101\rightarrow 1100\rightarrow 1000\rightarrow 0000)$ 
 \item $\pi_7$: $(0000\rightarrow 0100\rightarrow 0110\rightarrow 0111\rightarrow 1111\rightarrow 0111\rightarrow 0110\rightarrow 0100\rightarrow 0000)$ 
 \item $\pi_8$: $(0000\rightarrow 1000\rightarrow 1010\rightarrow 1011\rightarrow 1111\rightarrow 1011\rightarrow 1010\rightarrow 1000\rightarrow 0000)$ 
\end{itemize}

\begin{figure}[!ht]
\centering
\subfigure[Voltage $v_{1} = f(t)$]{ \includegraphics[scale = 0.30]{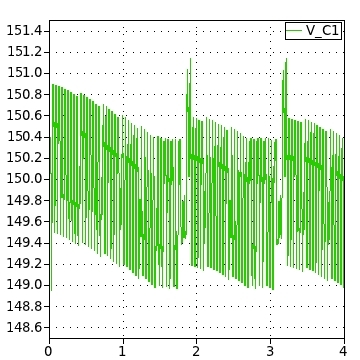}} 
\subfigure[Voltage $v_{2} = f(t)$]{ \includegraphics[scale = 0.30]{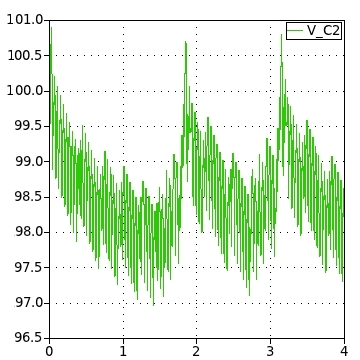}} 
\subfigure[Voltage $v_{3} = f(t)$]{ \includegraphics[scale = 0.30]{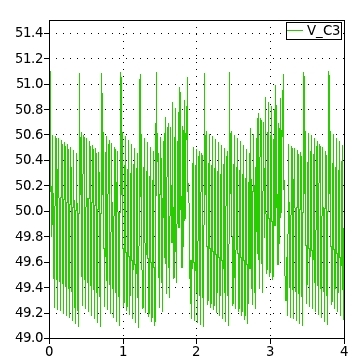}} \\
\subfigure[Projection in plane $(v_1,v_{2})$]{ \includegraphics[scale = 0.30]{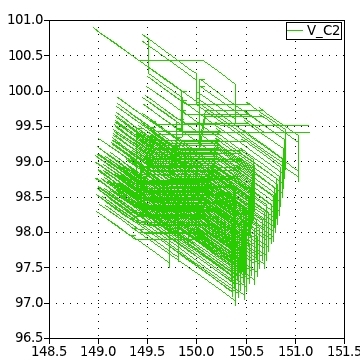}} 
\subfigure[Projection in plane $(v_1,v_{3})$]{ \includegraphics[scale = 0.30]{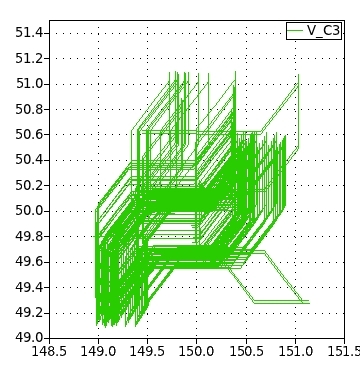}} 
\subfigure[Projection in plane $(v_{2},v_3)$]{ \includegraphics[scale = 0.30]{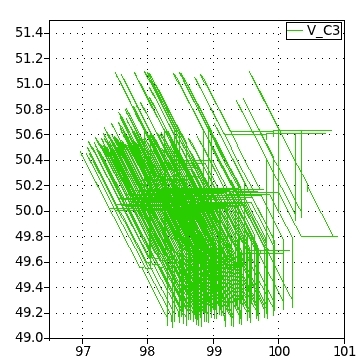}} \\
\caption{Capacitor voltages}
\label{fig:capacitor_voltages}`
\end{figure}

\begin{figure}[!ht]
 \centering
\subfigure[Current $i$]{ \includegraphics[scale = 0.35]{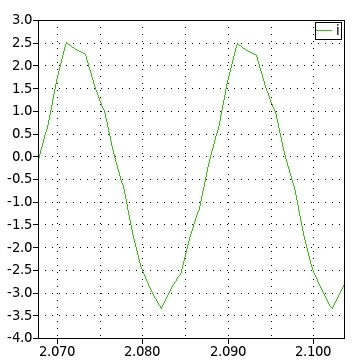}} 
\subfigure[Output voltage $v_o$]{ \includegraphics[scale = 0.20]{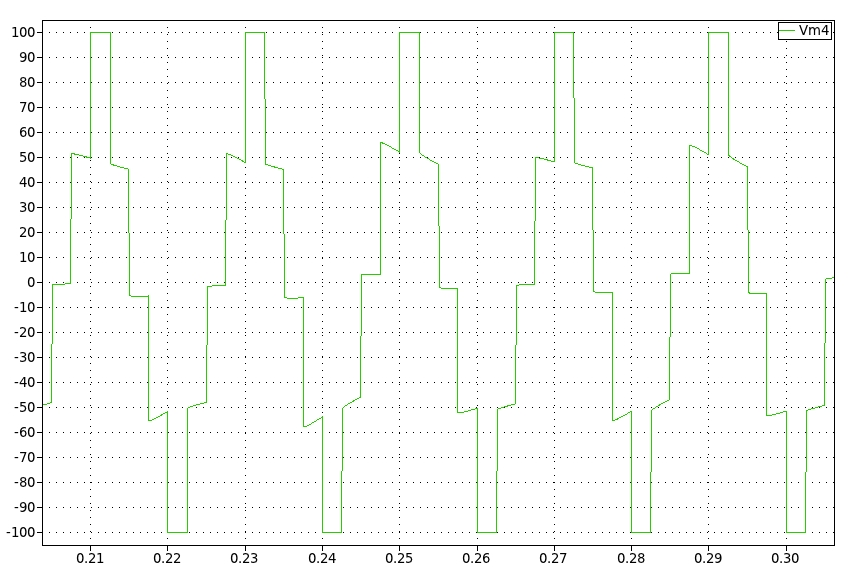}} \\
\caption{Current and output voltage}
\label{fig:current_output}
\end{figure}

By Proposition \ref{prop:safe}, the control of $R$ induced by $\Delta$ is safe w.r.t. $S$:
under the control induced by $\Delta$, all the trajectories starting from $R$ always stay in $S$. 
This guarantees that the property of capacitor voltage balance
is satisfied.
We present in Figures \ref{fig:capacitor_voltages} and 
\ref{fig:current_output}
a numerical simulation of this controller on the system starting from the point $v_{1}(0) = 150V, v_{2}(0) = 100V, v_{3}(0) = 50V \ \mathrm{and} \ i(0) = -3A$.
This simulation has been performed using tool PLECS \cite{plecs-web}.
One can check on the simulation that the system state always stays inside $S$.

\subsection{Application of the Decomposition procedure to a 7-level converter}\label{sec:7-level}
We now consider the case
of an $\ell$-level converter with $\ell=7$.
There are now 6 pairs of switching cells and 5 capacitors $C_1,\dots, C_5$.
The state of the system is $x(t)=[v_1(t)\ v_2(t)\ v_3(t)\ v_4(t)\ v_5(t)\ i(t)]^T$
where $v_j(t)$ is the voltage across $C_j$ ($1\leq j\leq 5$) and
$i(t)$ is the current flowing in the circuit.
The generated waveform now goes from $-v_{input}$ up to $+v_{input}$ 
with steps at $-\frac{2}{3}v_{input}$,$-\frac{1}{3}v_{input}$, $0$,
$\frac{1}{3}v_{input}$,$\frac{2}{3}v_{input}$,
and the cycle duration is $T=12\tau$.
There are now $518,400$  possible 
patterns for generating an $7$-level staircase signal on 1 cycle. 
We used the following values for the system constants:
output at $50 \mathrm{Hz}$,\footnote{which corresponds to $T=12\tau=0.02s$}
capacitances of $0.1F$,
resistor values $50 \Omega$,
inductor values $0.137 H$,
$v_{input} = 300V$.
Ideally, the output is thus
a staircase waveform with an amplitude of $600$V, centered around $0$V,
and the ideal values $v_i^*$ of the capacitor voltages 
of the capacitor $C_i$ are given by:
$v_1^* = 500V$, $v_2^*=400V$, $v_3^* = 300V$, $v_4^* = 200V$, $v_5^* = 100V$.
For $R$, we take
$R = [495,505] \times [395,405] \times [295,305] \times [195,205] \times [95,105]$,
which corresponds to a product of intervals centered around the
ideal values with a variation of $\pm 5V$ (i.e., $5\%$ 
of the least charged capacitor $C_5$).
For $S$, we take $R+\varepsilon$ with $\varepsilon=1V$,
which means that we have an additional tolerance of $\pm 1V$ for the fluctuations
occurring between two beginnings of cycle.
On the same machine as in Section \ref{sec:5-level}, 
with parameters $d=1$ and $k=12$, MINIMATOR outputs 
in 98 minutes
a decomposition~$\Delta$ which is safe w.r.t. $S$. See \cite{rr-lsv-12-16} for more details.
%

\begin{figure}[!ht]
\centering
\includegraphics[scale = 0.60]{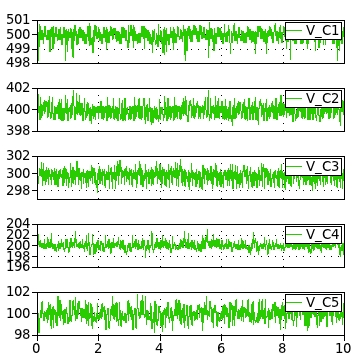}
\caption{Capacitor voltages}
\label{fig:7-capacitor_voltages}
\end{figure}

\begin{figure}[!ht]
 \centering
\subfigure[Current $i$]{ \includegraphics[scale = 0.3]{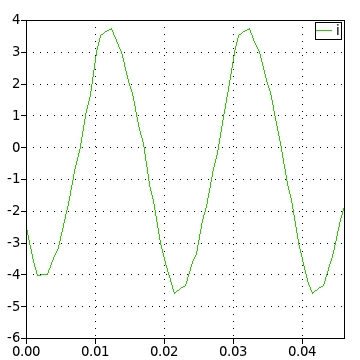}} 
\subfigure[Voltage $v_o$]{ \includegraphics[scale = 0.3]{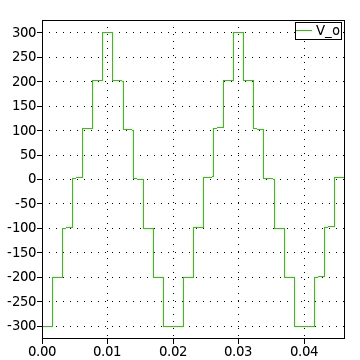}} \\
\caption{Current and output voltage}
\label{fig:7-current_output}
\end{figure}

We present in Figures
\ref{fig:7-capacitor_voltages}
and
\ref{fig:7-current_output}
a numerical simulation of the controlled system starting from the point $v_{1}(0) = 500V$, $v_{2}(0) = 400V$, $v_{3}(0) = 300V$, $v_{4}(0) = 200V$, $v_{5}(0) = 100V$ and $i(0) = -2.5A$.
One can check again on the simulation that the system state always stays inside $S$.

It is difficult to perform experiments with $\ell$ greater than 7
with the present implementation. The complexity
of the state decomposition procedure is 
indeed exponential in the number $\ell$ of levels.
We are presently implementing MINIMATOR on a parallel computing architecture
(see \cite{minimator})
in order to increase the tractable number of levels.

\section{Physical Experimentations on a 5-level Converter}\label{sec:physical_exp}
\begin{figure}[!ht]
 \centering
 \includegraphics[scale = 0.3]{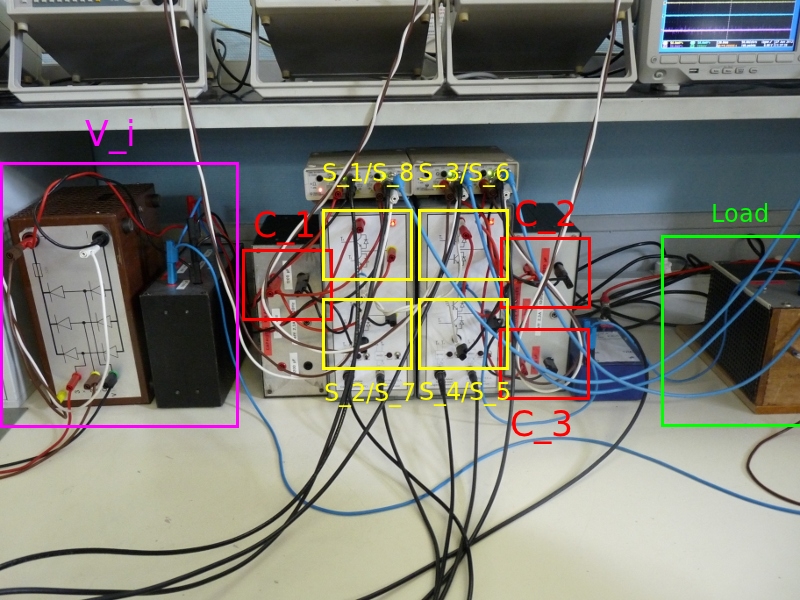}
 \caption{Prototype built by SATIE}
 \label{fig:photo-montage}
\end{figure}

\begin{figure}[!ht]
 \centering
 \includegraphics[scale = 0.3]{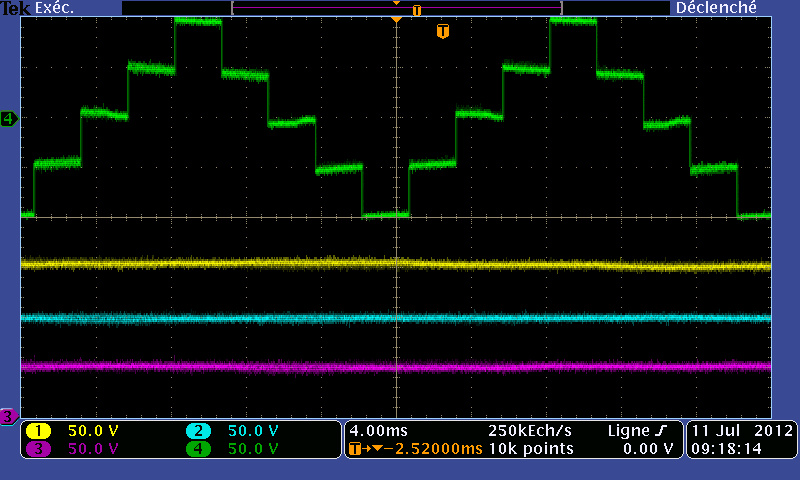}
 \caption{Output voltage (above in green) and capacitor voltages (below)}
 \label{fig:results-satie}
\end{figure}

\begin{figure}[!ht]
 \centering
 \includegraphics[scale = 0.3]{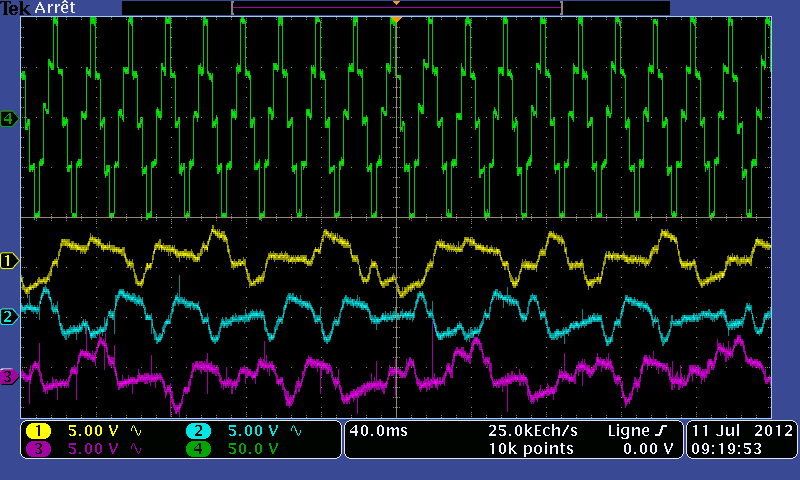}
 \caption{Zoom of output voltage (above) and capacitors voltages (below)}
 \label{fig:results-satie2}
\end{figure}

\begin{figure}[!ht]
 \centering
 \includegraphics[scale = 0.3]{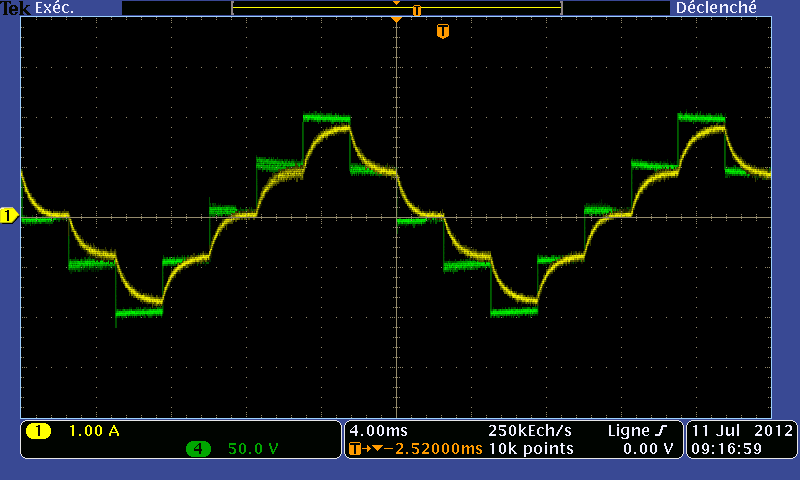}
 \caption{Output voltage (in green) and current (in yellow, after appropriate resizing) in the circuit}
 \label{fig:results-satie3}
\end{figure}

\begin{figure}[!ht]
 \centering
 \includegraphics[scale = 0.3]{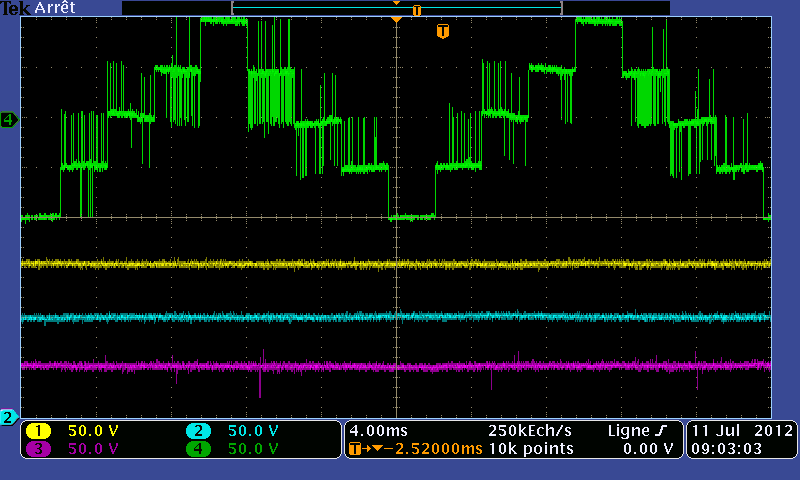}
 \caption{Output voltage (above) and capacitor voltages (below)
in presence of time-varying period $T$}
 \label{fig:results-satie4}
\end{figure}

A prototype
of the 5-level flying capacitor has been realized by the SATIE Laboratory 
in order to test our control strategy on an actual system. 
See Figure \ref{fig:photo-montage} for a picture of the prototype.
Our control strategy was applied to the system via Simulink and a dSpace\textsuperscript{\textregistered} interface.
The results are presented in Figure \ref{fig:results-satie} for the output voltage and the capacitor charges. In Figure \ref{fig:results-satie2}, we present the same results but with a larger scale on the capacitor voltage to see the fluctuations around the reference values.
As we can see, the experimental results 
are very closed to those obtained by simulation with PLECS of Section \ref{sec:5-level}.
In Figure~\ref{fig:results-satie3}, we represent the output voltage 
 together with the current (after appropriate
resizing) flowing the load.
During the experimentations, we have successfully 
tested the robustness of the controller
in presence of the following perturbations:
\begin{enumerate}
 \item The ideal voltage source as input is no longer ideal but its values fluctuate around the reference value.
 \item We use a time-varying period $T$ of cycle (instead of a constant one),
and check the preservation of the capacitor voltages balance.
The result of this experiment is depicted in Figure \ref{fig:results-satie4}. 
\end{enumerate}

Although these preliminary tests of robustness are promising,
they need to be consolidated, in particular in presence of
significant variations of resistor loads. 

\section{Final Remarks}\label{sec:conclusion}
We have synthesized a control strategy for a $5$-level 
and a 7-level flying capacitor converters using the method of state space decomposition.
This control is state-dependent and is interesting because:
\begin{itemize}
 \item at each electrical cycle, the controller indicates all the subsequent switching modes needed to produce one
period of the output voltage (instead of just the next switching mode),
 \item the controller takes into account only the capacitor voltages state and not the intensity state;
this is interesting because 
for practical applications, a current sensor is not always desired
(see \cite{DTOC09}). 
\end{itemize}
 We have checked by numerical simulations and physical experimentations that the control satisfies the capacitor voltage balancing
 and the staircase shape of 
the output voltage. We have also checked the robustness of the method with respect to several sources of perturbation. 

The method can be easily refined in order to generate
sinusoidal-like output signals rather than 
the triangular-like output signals generated here:
it suffices to adjust the switching instants within the  period  $T$
of the cycle, instead of using uniformly $\tau$. 

The method 
can be applied in principle to any number of levels for the flying capacitor
topology. However, it suffers from an exponential 
increase of complexity
when the level $\ell$ grows:
the method reaches its limit for $\ell=9$, which corresponds 
to a dimension $n=7$ of the state space.

For $\ell=5,7$,
the Decomposition procedure is well-suited to the {\em flying capacitor}
topology: the pattern length input $k$ is $2\times({\ell-1})$ where $\ell$ is the number of levels
of the converter, and the depth input $d$ is 1,
which means that the decomposition is found after a single bisection.
Note however that such simple decompositions of the state space
do not necessarily exist
for other topologies:
for multilevel modular converter toplogy \cite{LM03},
we had to propose in \cite{SHLRFLF-epe13}
a different and specialized algorithm which
takes additionally into account the value of the intensity state.

In future work, we plan to improve the robustness of 
the decomposition method for flying capacitor topology under variations of
the resistive and inductive load. This will allow us
to model the time-varying load of electrical
networks, which is a basic feature of electricity distribution,
and a major challenge today for renewable-energy technologies.
We are also implementing the tool MINIMATOR on a parallel computing 
architecture in order to synthesize correct-by-design controls
for multilevel converters with a greater numbe $\ell$ of levels.
\\

{\bf Acknowledgement.} We are grateful to St\'ephane Lefebvre for numerous helpful discussions. We also thank the anonymous referees for their constructive 
comments.
This work has been done within the framework of projects BOOST and BOOST2 supported by Institut Farman.
\bibliographystyle{eptcs}
\bibliography{biblio_bis}

\begin{thebibliography}{10}
\providecommand{\bibitemdeclare}[2]{}
\providecommand{\surnamestart}{}
\providecommand{\surnameend}{}
\providecommand{\urlprefix}{Available at }
\providecommand{\url}[1]{\texttt{#1}}
\providecommand{\href}[2]{\texttt{#2}}
\providecommand{\urlalt}[2]{\href{#1}{#2}}
\providecommand{\doi}[1]{doi:\urlalt{http://dx.doi.org/#1}{#1}}
\providecommand{\bibinfo}[2]{#2}

\bibitemdeclare{inproceedings}{CPM09}
\bibitem{CPM09}
\bibinfo{author}{I.~\surnamestart Cervantes\surnameend}, \bibinfo{author}{F.J.
  \surnamestart Perez-Pinal\surnameend} \& \bibinfo{author}{A.~\surnamestart
  Mendoza-Torres\surnameend} (\bibinfo{year}{2009}):
  \emph{\bibinfo{title}{Hybrid {C}ontrol of {DC}-{DC} {P}ower {C}onverters}}.
\newblock In: {\sl \bibinfo{booktitle}{Renewable {E}nergy (Chapter 10)}},
  \bibinfo{publisher}{T J Hammons}, pp. \bibinfo{pages}{173,193},
  \doi{10.5772/7370}.

\bibitemdeclare{article}{DTOC09}
\bibitem{DTOC09}
\bibinfo{author}{Zhong \surnamestart Du\surnameend}, \bibinfo{author}{L.M.
  \surnamestart Tolbert\surnameend}, \bibinfo{author}{B.~\surnamestart
  Ozpineci\surnameend} \& \bibinfo{author}{J.N. \surnamestart
  Chiasson\surnameend} (\bibinfo{year}{2009}):
  \emph{\bibinfo{title}{Fundamental Frequency Switching Strategies of a
  Seven-Level Hybrid Cascaded H-Bridge Multilevel Inverter}}.
\newblock {\sl \bibinfo{journal}{IEEE Transactions on Power Electronics}}
  \bibinfo{volume}{24}(\bibinfo{number}{1}), pp. \bibinfo{pages}{25--33},
  \doi{10.1109/TPEL.2008.2006678}.

\bibitemdeclare{techreport}{rr-lsv-12-16}
\bibitem{rr-lsv-12-16}
\bibinfo{author}{G.~\surnamestart Feld\surnameend},
  \bibinfo{author}{L.~\surnamestart Fribourg\surnameend},
  \bibinfo{author}{D.~\surnamestart Labrousse\surnameend},
  \bibinfo{author}{B.~\surnamestart Revol\surnameend} \&
  \bibinfo{author}{R.~\surnamestart Soulat\surnameend} (\bibinfo{year}{2012}):
  \emph{\bibinfo{title}{Correct by design control of 5-level and 7-level
  converters}}.
\newblock \bibinfo{type}{Research Report} \bibinfo{number}{LSV-12-25},
  \bibinfo{institution}{Laboratoire Sp{\'e}cification et V{\'e}rification, ENS
  Cachan, France}.

\bibitemdeclare{inproceedings}{FS-rp13}
\bibitem{FS-rp13}
\bibinfo{author}{Laurent \surnamestart Fribourg\surnameend} \&
  \bibinfo{author}{Romain \surnamestart Soulat\surnameend}
  (\bibinfo{year}{2013}): \emph{\bibinfo{title}{Stability Controllers for
  Sampled Switched Systems}}.
\newblock In \bibinfo{editor}{Parosh~Aziz \surnamestart Abdulla\surnameend} \&
  \bibinfo{editor}{Igor \surnamestart Potapov\surnameend}, editors: {\sl
  \bibinfo{booktitle}{{P}roceedings of the 7th {W}orkshop on {R}eachability
  {P}roblems in {C}omputational {M}odels ({RP}'13)}}, {\sl
  \bibinfo{series}{Lecture Notes in Computer Science}} \bibinfo{volume}{8169},
  \bibinfo{publisher}{Springer}, \bibinfo{address}{Uppsala, Sweden}, pp.
  \bibinfo{pages}{135--145}, \doi{10.1007/978-3-642-41036-9\_13}.

\bibitemdeclare{article}{GPT10}
\bibitem{GPT10}
\bibinfo{author}{A.~\surnamestart Girard\surnameend},
  \bibinfo{author}{G.~\surnamestart Pola\surnameend} \&
  \bibinfo{author}{P.~\surnamestart Tabuada\surnameend} (\bibinfo{year}{2010}):
  \emph{\bibinfo{title}{Approximately Bisimilar Symbolic Models for
  Incrementally Stable Switched Systems}}.
\newblock {\sl \bibinfo{journal}{IEEE Trans. on Automatic Control}}
  \bibinfo{volume}{55}, pp. \bibinfo{pages}{116--126},
  \doi{10.1109/TAC.2009.2034922}.

\bibitemdeclare{article}{K98}
\bibitem{K98}
\bibinfo{author}{W.~\surnamestart K\"{u}hn\surnameend} (\bibinfo{year}{1998}):
  \emph{\bibinfo{title}{Zonotope dynamics in numerical quality control}}.
\newblock {\sl \bibinfo{journal}{Mathematical {V}isualization}}, pp.
  \bibinfo{pages}{125--134}, \doi{10.1007/978-3-662-03567-2\_10}.

\bibitemdeclare{inproceedings}{LM03}
\bibitem{LM03}
\bibinfo{author}{A.~\surnamestart Lesnicar\surnameend} \&
  \bibinfo{author}{R.~\surnamestart Marquardt\surnameend}
  (\bibinfo{year}{2003}): \emph{\bibinfo{title}{An innovative modular
  multilevel converter topology suitable for a wide power range}}.
\newblock In: {\sl \bibinfo{booktitle}{Power Tech Conference Proceedings, 2003
  IEEE Bologna}}, \bibinfo{volume}{3}, pp. \bibinfo{pages}{6 pp. Vol.3--},
  \doi{10.1109/PTC.2003.1304403}.

\bibitemdeclare{inproceedings}{MF92}
\bibitem{MF92}
\bibinfo{author}{T.A. \surnamestart Meynard\surnameend} \&
  \bibinfo{author}{H.~\surnamestart Foch\surnameend} (\bibinfo{year}{1992}):
  \emph{\bibinfo{title}{Multi-level conversion: high voltage choppers and
  voltage-source inverters}}.
\newblock In: {\sl \bibinfo{booktitle}{23rd Annual IEEE Power Electronics
  Specialists Conference (PESC '92)}}, \bibinfo{volume}{1}, pp.
  \bibinfo{pages}{397--403}, \doi{10.1109/PESC.1992.254717}.

\bibitemdeclare{misc}{minimator}
\bibitem{minimator}
\emph{\bibinfo{title}{{MINIMATOR} {W}eb Page}}.
\newblock \bibinfo{howpublished}{{https://bitbucket.org/ukuehne/minimator/}}.

\bibitemdeclare{misc}{octave}
\bibitem{octave}
\emph{\bibinfo{title}{{Octave} {W}eb Page}}.
\newblock \bibinfo{howpublished}{{http://www.gnu.org/software/octave/}}.

\bibitemdeclare{misc}{plecs-web}
\bibitem{plecs-web}
\emph{\bibinfo{title}{{PLECS} {W}eb Page}}.
\newblock \bibinfo{howpublished}{{http://www.plexim.com}}.

\bibitemdeclare{article}{SMVSSDSB12}
\bibitem{SMVSSDSB12}
\bibinfo{author}{B.~\surnamestart Singh\surnameend},
  \bibinfo{author}{N.~\surnamestart Mittal\surnameend}, \bibinfo{author}{K.S.
  \surnamestart Verma\surnameend}, \bibinfo{author}{D.~\surnamestart
  Singh\surnameend}, \bibinfo{author}{S.P. \surnamestart Singh\surnameend},
  \bibinfo{author}{R.~\surnamestart Dixit\surnameend},
  \bibinfo{author}{M.~\surnamestart Singh\surnameend} \&
  \bibinfo{author}{A.~\surnamestart Baranwal\surnameend}
  (\bibinfo{year}{2012}): \emph{\bibinfo{title}{Multi-level inverter: A
  literature survey on topologies and control strategies}}.
\newblock {\sl \bibinfo{journal}{International Journal of Reviews in
  Computing}} \bibinfo{volume}{10}.

\bibitemdeclare{inproceedings}{SHLRFLF-epe13}
\bibitem{SHLRFLF-epe13}
\bibinfo{author}{Romain \surnamestart Soulat\surnameend},
  \bibinfo{author}{Guillaume \surnamestart H{\'e}rault\surnameend},
  \bibinfo{author}{Denis \surnamestart Labrousse\surnameend},
  \bibinfo{author}{Bertrand \surnamestart Revol\surnameend},
  \bibinfo{author}{Gilles \surnamestart Feld\surnameend},
  \bibinfo{author}{St{\'e}phane \surnamestart Lefebvre\surnameend} \&
  \bibinfo{author}{Laurent \surnamestart Fribourg\surnameend}
  (\bibinfo{year}{2013}): \emph{\bibinfo{title}{Use of a full wave
  correct-by-design command to control a multilevel modular converter}}.
\newblock In \bibinfo{editor}{{\relax Ph}ilippe \surnamestart
  Lataire\surnameend}, editor: {\sl \bibinfo{booktitle}{{P}roceedings of the
  15th {E}uropean {C}onference on {P}ower {E}lectronics and {A}pplications
  ({EPE}'13)}}, \bibinfo{publisher}{{IEEE} Power Electronics Society},
  \bibinfo{address}{Lille, France}, pp. \bibinfo{pages}{1,8},
  \doi{10.1109/EPE.2013.6634448}.
\newblock
  \urlprefix\url{http://www.lsv.ens-cachan.fr/Publis/PAPERS/PDF/SHLRFLF-epe13.%
pdf}.

\bibitemdeclare{book}{T09}
\bibitem{T09}
\bibinfo{author}{Paulo \surnamestart Tabuada\surnameend}
  (\bibinfo{year}{2009}): \emph{\bibinfo{title}{Verification and Control of
  Hybrid Systems: A Symbolic Approach}}.
\newblock \bibinfo{publisher}{Springer Publishing Company, Incorporated},
  \doi{10.1007/978-1-4419-0224-5}.

\end{thebibliography}

\end{document}